\documentclass[conference]{IEEEtran}%
\usepackage[cmex10]{amsmath}
\usepackage[cmex10]{amsmath}

\usepackage{ifpdf}
\usepackage{algorithm}
\usepackage{algpseudocode}
\usepackage{xcolor}
\usepackage{url}
\ifpdf
\usepackage[pdftex]{graphicx}
\else
\usepackage[dvips]{graphicx}
\fi
\usepackage{tabularx}
\usepackage{multirow}
\graphicspath{{graphics/png/}}
\DeclareGraphicsExtensions{.png, .eps, .pdf}
\usepackage{float} 
\usepackage{subfigure}
\usepackage{comment}
\usepackage{bm}
\usepackage{amsmath,amsthm,amsfonts,amssymb}
\usepackage[noadjust]{cite}
\usepackage{nomencl}
\usepackage{hyperref}
\theoremstyle{definition}

\newcommand\Tstrut{\rule{0pt}{2.6ex}}         

\newcommand{\done}[1]{\textcolor{black}{#1}}
\newcommand{\rev}[1]{\textcolor{black}{#1}}

\IEEEoverridecommandlockouts
\begin{document}
%
\title{Energy Storage Price Arbitrage via \\ Opportunity Value Function Prediction}


\author{Ningkun Zheng\IEEEauthorrefmark{1}, \IEEEauthorblockN{Xiaoxiang Liu\IEEEauthorrefmark{2}, Bolun Xu\IEEEauthorrefmark{1}}
\IEEEauthorblockA{\IEEEauthorrefmark{1}\textit{Earth and Environmental Engineering}, \IEEEauthorrefmark{2}\textit{Computer Science} \\
\textit{Columbia University}\\
New York, New York, US \\
\{nz2343,xl2948,  bx2177\}@columbia.edu }
\and
\IEEEauthorblockN{Yuanyuan Shi}
\IEEEauthorblockA{\textit{Electrical and Computer Engineering} \\
\textit{University of California San Diego}\\
San Diego, California, US \\
yyshi@eng.ucsd.edu }

}


\maketitle

\begin{abstract}
This paper proposes a novel energy storage price arbitrage algorithm combining supervised learning with dynamic programming. The proposed approach uses a neural network to directly predicts the opportunity cost at different energy storage state-of-charge levels, and then input the predicted opportunity cost into a model-based arbitrage control algorithm for optimal decisions. We generate the historical optimal opportunity value function using price data and a dynamic programming algorithm, then use it as the ground truth and historical price as predictors to train the opportunity value function prediction model.
Our method achieves 65\% to 90\% profit compared to perfect foresight in case studies using different energy storage models and price data from New York State, which significantly outperforms existing model-based and learning-based methods. While guaranteeing high profitability, the algorithm is also light-weighted and can be trained and implemented with minimal computational cost.
Our results also show that the learned prediction model has excellent transferability. The prediction model trained using price data from one region also provides good arbitrage results when tested over other regions. 
\end{abstract}


\IEEEpeerreviewmaketitle

\vspace{-1em}
\section{Introduction}
With technological advances and a decarbonization target in 2030~\cite{room2021fact}, battery energy storage systems are becoming the cornerstone flexibility resource in future decarbonized power systems to balance electricity supply and demand~\cite{waite2020electricity, wang2017impact}.
As progressively installed energy storage saturates ancillary service markets~\cite{ma2021data}, energy storage participants increasingly focus on arbitraging in wholesale energy markets~\cite{eia2022}. Energy storage arbitrage by charging during low price periods and discharging during high price periods, earning revenues while aiding power system operations based on price signals.

Real-time electricity prices are highly volatile and stochastic, and participants must combine the energy storage physical constraints with the price models to make arbitrage decisions. Model-predictive control (MPC) is the most common approach for price arbitrage. MPC approaches~\cite{tesla_arbitrage, arnold2011model} follow a two-step process that predicts the price time series first, followed by model-based optimizations.
The precise prediction of real-time price series is thus critical. 
Various learning methods have been implemented to predict real-time prices, including locally linear neuro-fuzzy model~\cite{sarafraz2011locational}, support vector machine~\cite{nwulu2013soft}, and residual neural network~\cite{chaweewat2020electricity}. However, most of these methods focus on price prediction over a short look-ahead window, such as one or two hours. The prediction accuracy drops significantly at a longer prediction horizon, such as 24 hours, to capture the daily price variations, which are essential for energy storage arbitrage~\cite{xie2012fast,zhang2020mpc,zheng2022comparing}. \rev{Solving iterative optimization problems also proposes excessive computation costs, which could limit the potential of MPC utensils in edge applications.} 

Alternative approaches for arbitrage include reinforcement learning (RL), stochastic dynamic programming (SDP), and stochastic dual dynamic programming (SDDP). SDP and RL approaches share similar challenges in the computation time as it requires discretization of action, state, and prices to construct the Markov-decision process (MDP). Especially, arbitrage in the real-time market requires including finer states and actions in MDP due to shorter market periods~\cite{wang2018energy,cao2020deep}. Also, modeling non-perfect efficiencies in MDP is also challenging as the state transition logic becomes much more sophisticated. 
\rev{In contrast, SDDP does not discretize state and action spaces but requires forward-backward iterations to generate dual cuts~\cite{megel2015stochastic}, and also requires hand-tuning price distribution as Markov processes. 
}

In this paper, we propose a novel approach that achieves state-of-the-art arbitrage results while the training process is computationally and data efficient. \rev{Our contribution includes:}
\begin{enumerate}
    \item Motivated by dynamic programming, we combine supervised learning with model-based optimization into a novel approach to directly predict the opportunity cost function of energy storage state-of-charge (SoC) \done{and further use it in arbitrage control}.
    \item Our approach significantly reduces the training load while providing reliable arbitrage decisions which \done{leverage costs and} satisfy physical constraints. The proposed model has stable performance with minimal hyper-parameters, and the arbitrage profit surpass benchmarks with RL and SDP approach with significantly less training time.
    \item We test the proposed algorithm using real price data from four New York ISO price zones. The result shows the model achieved up to 90\% of profit compared to perfect foresight with varying durations and costs. The result also shows our model has excellent transferability and still achieves great profit results when trained and tested using price data from different locations.
\end{enumerate}

\rev{The rest of the paper is organized as follows. Sections~\ref{sec:formulation} and \ref{sec:solution} present the optimization problem, solution and training algorithm for energy storage arbitrage. Section~\ref{sec:result} illustrate our algorithm with real data numerical simulations. Section~\ref{sec:conclusion} concludes this paper.}

\section{Problem Definition and Formulation}\label{sec:formulation}
We consider a price response problem in which the energy storage can decide whether to charge or discharge immediately after observing a new price signal.
Our approach is to predict the opportunity \done{value (cost)} at different energy storage SoC levels and use the predicted values to optimize arbitrage decisions using observed prices and the energy storage model. We will present the arbitrage formulation and then define the learning problem to predict the \done{opportunity value} in this section.

\subsection{Arbitrage Formulation with \done{Opportunity Value} Prediction}
Motivated by dynamic programming~\cite{xu2020operational}, we formulate the arbitrage problem as an iterative single-period maximizing problem
\begin{subequations}\label{eq:p1}
\begin{align}
    \textstyle\max_{ \substack{b_t, p_t, e_t \\ \in \mathcal{E}(e_{t-1})}} \lambda_t (p_t-b_t) - cp_t + \hat{V}\big(e_{t}|\bm{\theta}, \bm{x}\big). 
    \label{eq:obj1}
\end{align}
where the first term represents the arbitrage revenue, which is the product of the real-time market price $\lambda_t$ and the energy storage dispatch decision $(p_t-b_t)$, $p_t$ is the discharge output, and $b_t$ is the charge output. The second term models operation and maintenance costs to prevent frequent cycling, where $c$ is the marginal discharge cost. The third term $\hat{V}\big(e_{t}|\bm{\theta}, \bm{x}\big)$ is the predicted energy storage opportunity value\rev{, which is a function of the final SoC $e_t$ at the end of time period $t$}. $\hat{V}$ depends on the prediction model parameters $\bm{\theta}$ and the prediction features $\bm{x}_{[t-1,t-2,\dotsc,t-W]}$ over a look-back period $W$.

We denote that the energy storage charge and discharge decisions and the final SoC belong to a feasibility set $\mathcal{E}(e_{t-1})$, which is dependent on the energy storage starting SoC $e_{t-1}$ at the start of time period $t$ (same as by the end of time period $t-1$). $\mathcal{E}(e_{t-1})$ is described with the following constraints:
\begin{gather}
    0 \leq b_t \leq P,\; 0\leq p_t \leq P \label{p1_c2} \\
    \text{$p_t = 0$ if $\lambda_t < 0$} \label{p1_c5}\\
    e_t - e_{t-1} = -p_t/\eta^p + b_t\eta^b \label{p1_c1}\\
    0 \leq e_t \leq E \label{p1_c3}
\end{gather}
\end{subequations}
where \eqref{p1_c2} models the upper and lower bound of charge and discharge output.
\done{\eqref{p1_c5} is a relaxed form of constraint that prevents simultaneous  discharge and charge, because the negative price is the necessary condition for energy storage to charge and discharge simultaneously in price arbitrage.}
\eqref{p1_c1} models the energy storage SoC evolution constraint with discharge and charge efficiency $\eta^p,\eta^b$. \eqref{p1_c3} models the energy storage SoC level upper bound $E$ and lower bound (we assume as 0).

\subsection{Piece-wise Linear Opportunity Value Function Learning}

The objective of the problem in this paper is to optimize the prediction model parameters $\bm{\theta}$ that maximize energy storage arbitrage profit over a set of training price data and energy storage physical parameters. Intuitively, this problem could be formulated as a bi-level problem. The upper level maximizes the total profit over the entire training time horizon. The lower level enforces a non-anticipatory decision-making process in which the energy storage dispatch decision only depends on the current price and the predicted opportunity value as in \eqref{eq:p1}. However, it will likely be computationally intractable.


\textbf{Problem Statement. }In this paper, we consider an alternative two-stage training approach in which we first generate the historical optimal opportunity value function $v_t(e)$ \done{for all SoC level $e$ constrained by~\eqref{p1_c3}}  and then train the learning model to predict the opportunity value function. Since we do not assume any particular functional form of the SoC opportunity value function \rev{$v_t(e)$}, we apply piece-wise linear approximation in which we describe \rev{$v_t(e)$} using fixed SoC segments. Each segment is associated with an \rev{opportunity value $V_t(e)$ corresponding to the integral of $v_t(e)$}. The \textbf{first stage} of the training process is thus to calculate the historically optimal opportunity value function $v_t(e)$ using the following equations
\begin{subequations}\label{eq3}
\begin{align}
    V_{t-1}(e_{t-1}) &= \textstyle \max_{\substack{b_t, p_t, e_t \\ \in \mathcal{E}(e_{t-1})}} \lambda_t (p_t-b_t) - cp_t + V_{t}(e_{t}) 
    \label{eq:obj3}\\
    v_{t}(e) &= \frac{\partial}{\partial e}V_t(e)
\end{align}

Note that different from \eqref{eq:p1}, \eqref{eq:obj3} is deterministic, and opportunity value $V_{t}$ is calculated based on historical price data instead of predicted using a learning model.
The following section will further discuss the generation of optimal opportunity value functions $v_t(e)$. The \textbf{second stage} trains a learning model to predict the opportunity value function $v_t(e)$ using predictors which are past prices
\begin{align}
\min_{\theta} \sum_{e\in\mathcal{S}}\Big|\Big|\hat{v}\big(e|\bm{\theta}, \bm{x}_{[t-1,t-2,\dotsc,t-W]}\big) -  v_{t}(e) \Big|\Big|^2_2\label{eq:obj4}
\end{align}
\end{subequations}

The learning model thus fits a piece-wise linear approximation of the historical optimal opportunity value function $v_t(e)$ based on the first-order derivative of the optimal value $V_t$, and $e$ is from the set of SoC segments $\mathcal{S}$. \rev{The prediction $\hat{v}$ is the predicted opportunity value function, and its integral is the predicted opportunity value $\hat{V}$ in (\ref{eq:obj1}).} 

\rev{We represent the energy storage opportunity value function $\hat{v}$ using piece-wise linear approximation because it has no closed form but is a 
monotonic decreasing derivative~\cite{xu2020operational}. It is corresponding to the diminishing value of energy stored, which is similar to a demand curve.} 
We evenly discretize $\hat{v}(e|\bm{\theta}, \bm{x})$ into segments from the lowest SoC level to the highest SoC level $E$. In each SoC segment we assume the opportunity value function $\hat{v}$ is a constant function. 


\section{Solution and Training Method}\label{sec:solution}

Our technical approach includes three efforts. First, we use a deterministic price arbitrage dynamic programming model to generate the optimal opportunity value function $v_t(e)$ based on historical price data. We then train a learning model using the historical optimal opportunity value function to predict the future optimal opportunity value functions. Finally, we test the learned model over future price datasets.

We first solve the dynamic programming problem \eqref{eq:obj3} subject to constraints \eqref{p1_c2}--\eqref{p1_c3}. We utilize the algorithm in our prior work~\cite{zheng2022arbitraging} to solve \eqref{eq:obj3} analytically and obtain piece-wise linear approximations of the value functions $v_t(e)$ for all time periods. With high computational efficiency, we can discretize $v_t(e)$ by equally spaced energy storage SoC level $e$ into small segments, which is far smaller than the power rating $P$. For any SoC level $e_t$, we can find the nearest segment and return the corresponding value.

\rev{We train standard multilayer perceptron (MLP) networks using the Adam algorithm to predict future opportunity value functions $\hat{v}_t(e)$ as shown in Algorithm~\ref{alg:1}. We use day-ahead prices (DAPs) and real-time prices (RTPs) with look-back window $W$ as predictors. Relying on Adam, we can iteratively update $\bm{\theta}$ based on the gradient of (\ref{eq:obj4}) with respect to $\bm{x}$.}  


\begin{algorithm}
  \caption{Value Function Prediction Model Training}
  \begin{algorithmic}[1]
    \State\textbf{Initialization:} Set random seed. Set hyper-parameters in Adam (Default). Set number of neurons based on input data and ReLu as activation function. Initialize $\bm{\theta}$.
    \State $i\gets 0$
      \While{not meet stop criteria}
        \For{$t \in [1,t]$}
            \State $\bm{x} \gets $ DAPs/RTPs with look-back window $W$
            \State $v_{t}(e)\gets $ historical opportunity value function at $t$
            \State \textbf{Update} $\bm{\theta}$ by minimizing Eq.~(\ref{eq:obj4}) using Adam
        \EndFor
        \State $i \gets i+1$
      \EndWhile
      \State \textbf{return} $\bm{\theta}$\Comment{Parameters of the prediction model}
  \end{algorithmic}
  \label{alg:1}
\end{algorithm}

\rev{After training the value function prediction model, we use the algorithm~\ref{alg:2} for arbitrage control. \rev{We can calculate the opportunity value $\hat{V}\big(e_{t}|\bm{\theta}, \bm{x}\big)$ by integrating the value function prediction $\hat{v}\big(e_{t}|\bm{\theta}, \bm{x}\big)$. 
In our implementation, we can solve (\ref{eq:p1}) effortlessly using value function prediction $\hat{v}$ and first-order conditions, since it is a single-period problem \cite{zheng2022arbitraging}}.}

\begin{algorithm}
  \caption{Arbitrage with Value Function Prediction}
  \begin{algorithmic}[1]
    \State\textbf{Initialization:}
    \State Set energy storage parameters $c, P, \eta^p, \eta^b, E $.
    \State Initialize $e_{t-1}\gets e_0$.
      \For{\texttt{$t\in [1,T]$}}
        \State Predict $\hat{v}\big(e_{t}|\bm{\theta}, \bm{x}\big)$
        \State Solve single-period optimization (\ref{eq:p1})
        \State \textbf{Return} $e_t, p_t, d_t$
      \EndFor
  \end{algorithmic}
\label{alg:2}
\end{algorithm}

\section{Case Study}\label{sec:result}

\begin{figure}[ht]%
	\centering
	\vspace{-1em}
		\includegraphics[trim = 10mm 0mm 24mm 0mm, clip, width = .9\columnwidth]{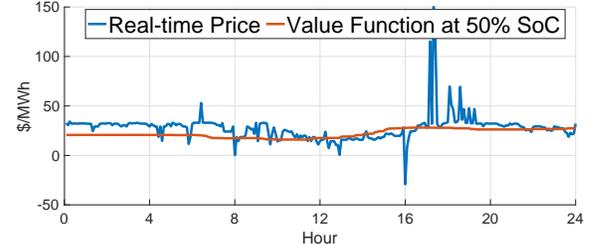}
    \vspace{-1em}
    \caption{Example of historical price data and optimal value function at 50\% SoC in NYC.}
    \label{fig:nyc}
\vspace{-0.6em}
\end{figure}

We pick four price zones in NYISO with different price patterns due to the significant transmission congestion within New York State~\cite{patton20162014}. We use RTPs and DAPs data from these zones with 5-minute and hourly resolutions, respectively. In all test cases, we use price data in 2017 and 2018 to calculate optimal opportunity value functions, then train the value function prediction model, and test the trained model by predicting value functions with price data in 2019 and performing arbitrage according to the control \rev{algorithm~\ref{alg:2}}. The example optimal value function at 50\% SoC and corresponding real-time price are shown in Fig~\ref{fig:nyc}. Compared to the optimal value function, the real-time price has more sparse price spikes and negative prices and, therefore, is hard to be predicted precisely by learning methods.

We use normalized energy storage capacity $E$ of 1~MWh, identical charge and discharge one-way efficiency $\eta$ of 90\%, marginal discharge cost $c$ of \$10/MWh, and power rating $P$ of 0.5~MW (2-hour duration), except specific stated. We generate the optimal value function for all time steps with 0.001~MWh granularity, thus 1001 SoC segments per function, and down-sample value functions to 50 segments. The down-sampled optimal value function $v_t$ is the ground truth of model training with \eqref{eq:obj4}. We use four MLP model settings with different input data look-back windows, model capacities, and stop criteria. Settings are shown in Table.~\ref{tab:setting}. Setting 1 takes the RTPs from the past three hours (36 price signals). Setting 2 extend RTPs look-back window to 24 hours (288 price signals). Whereas Setting 3 and 4 additionally include 24 hours DAPs as predictors. All settings have two hidden layer with same number of neurons.

\begin{table}[ht]
\footnotesize
\vspace{-1em}
\caption{MLP training input features and layers settings}
\centering
\vspace{-1em}
\resizebox{0.48\textwidth}{!}{
\begin{tabular}{ccccc}
\hline
\hline
\Tstrut
Setting & Num of RTP & Num of DAP   & Num of Neurons  & Num of Epochs\\
\hline
1&36&0&60&10\\
2&288&0&256&20\\
3&36&24&60&10\\
4&288&24&256&20\\
\hline
\hline
\end{tabular}
}
\label{tab:setting}
\vspace{-1em}
\end{table}


Computational costs with Google Colab (with GPU) are as follows\footnote{[Code Available]: \url{https://github.com/niklauskun/ES_NN_Arbitrage}}. For optimal value function generation, two year time horizon with 5-minute resolution (210240 time periods) can be solved within half minutes. For the training process, each epoch takes 8 seconds for settings Setting 1 and 3 and 12 seconds for settings Setting 2 and 4. Testing with one year data takes around 10 to 12 seconds (105120 time periods).

\subsection{Model Setting Comparison}

\begin{figure}[hb]
    \centering
    \vspace{-1.2em}
	\subfigtopskip=2pt
	\subfigbottomskip=2pt
	\subfigcapskip=-5pt
    \subfigure[NYC]{
		\includegraphics[trim = 0mm 0mm 10mm 0mm, clip, width = .45\columnwidth]{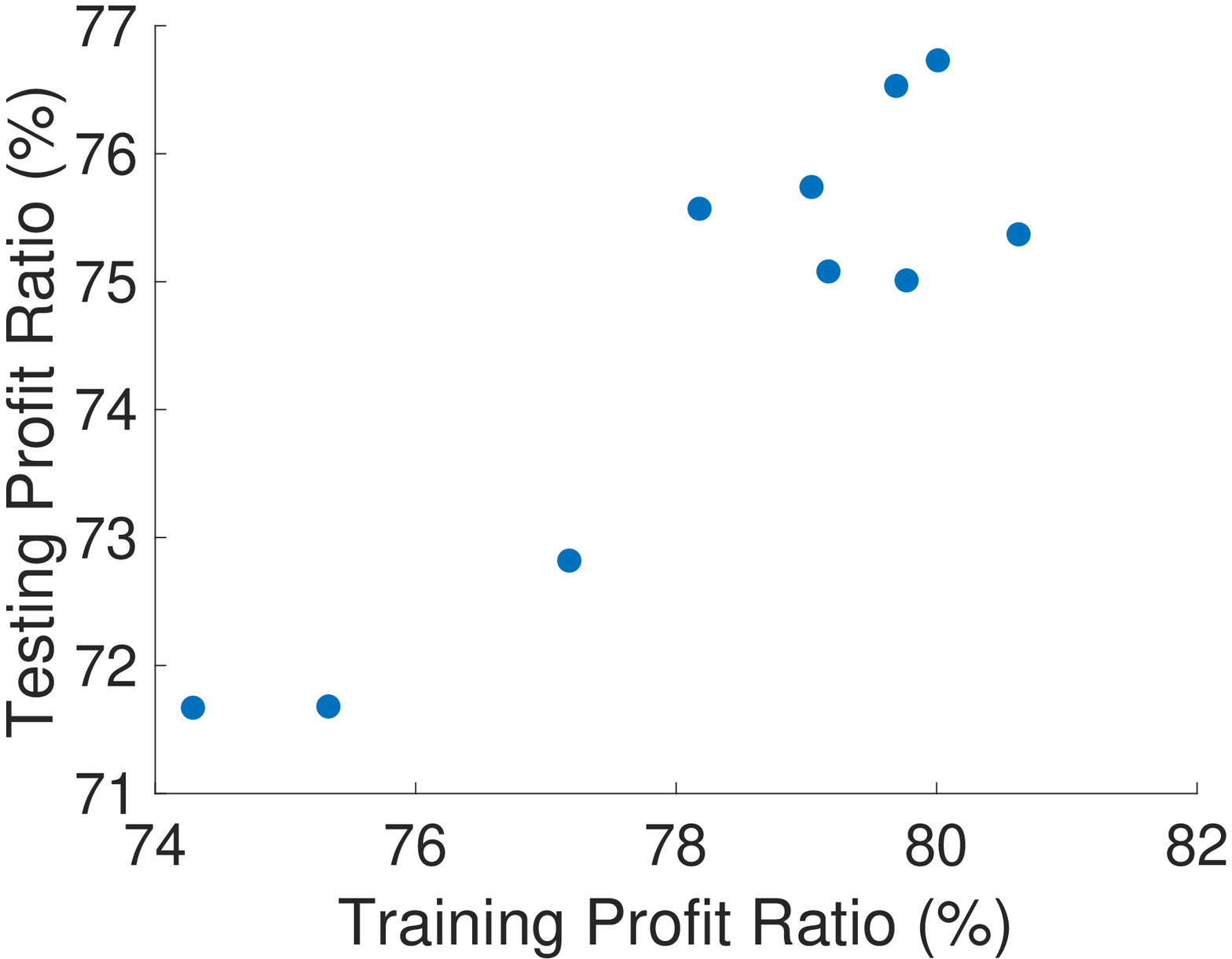}}
    \subfigure[LONGIL]{
		\includegraphics[trim = 0mm 0mm 10mm 0mm, clip, width = .45\columnwidth]{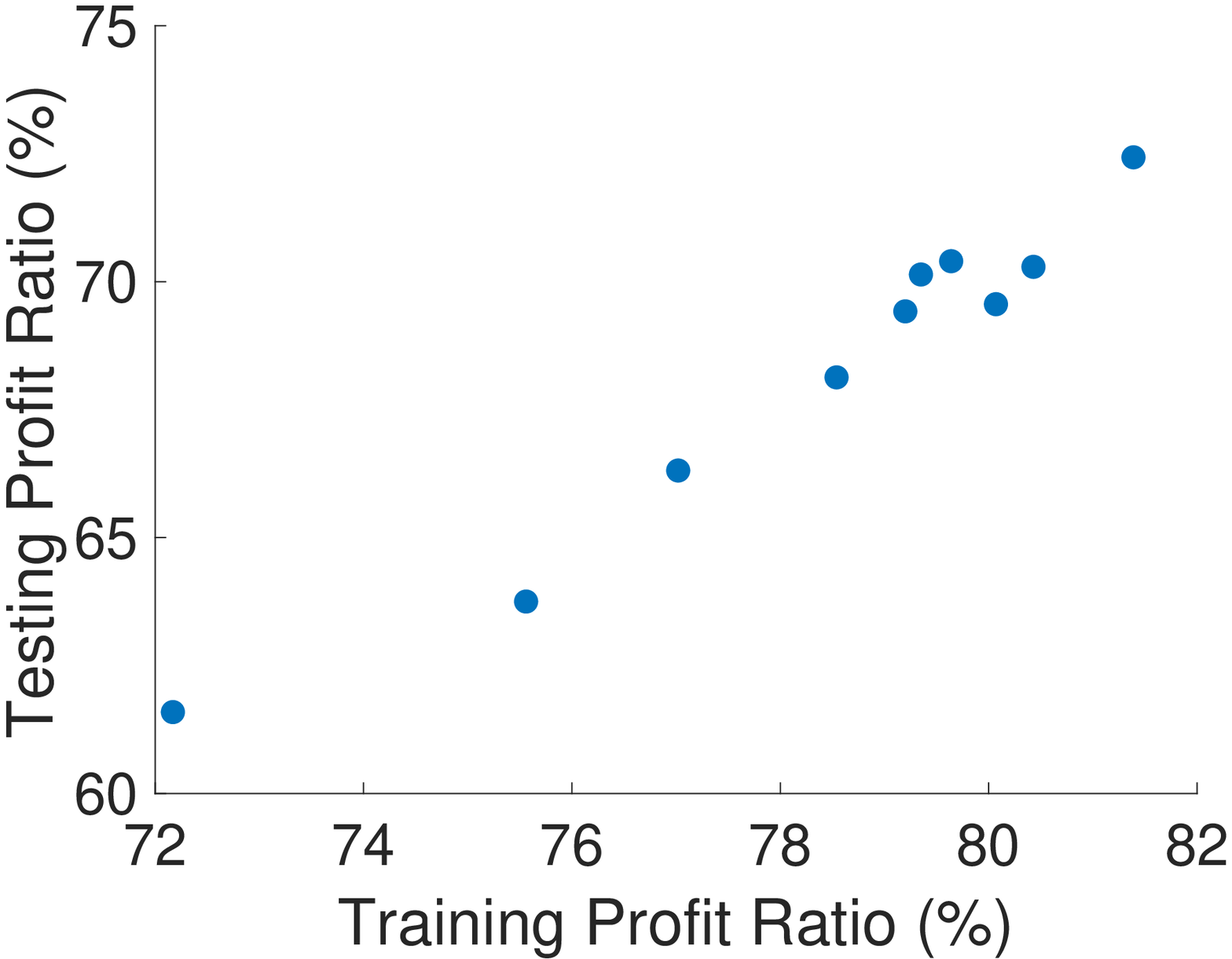}}
    \\
    \subfigure[NORTH]{
		\includegraphics[trim = 0mm 0mm 10mm 0mm, clip, width = .45\columnwidth]{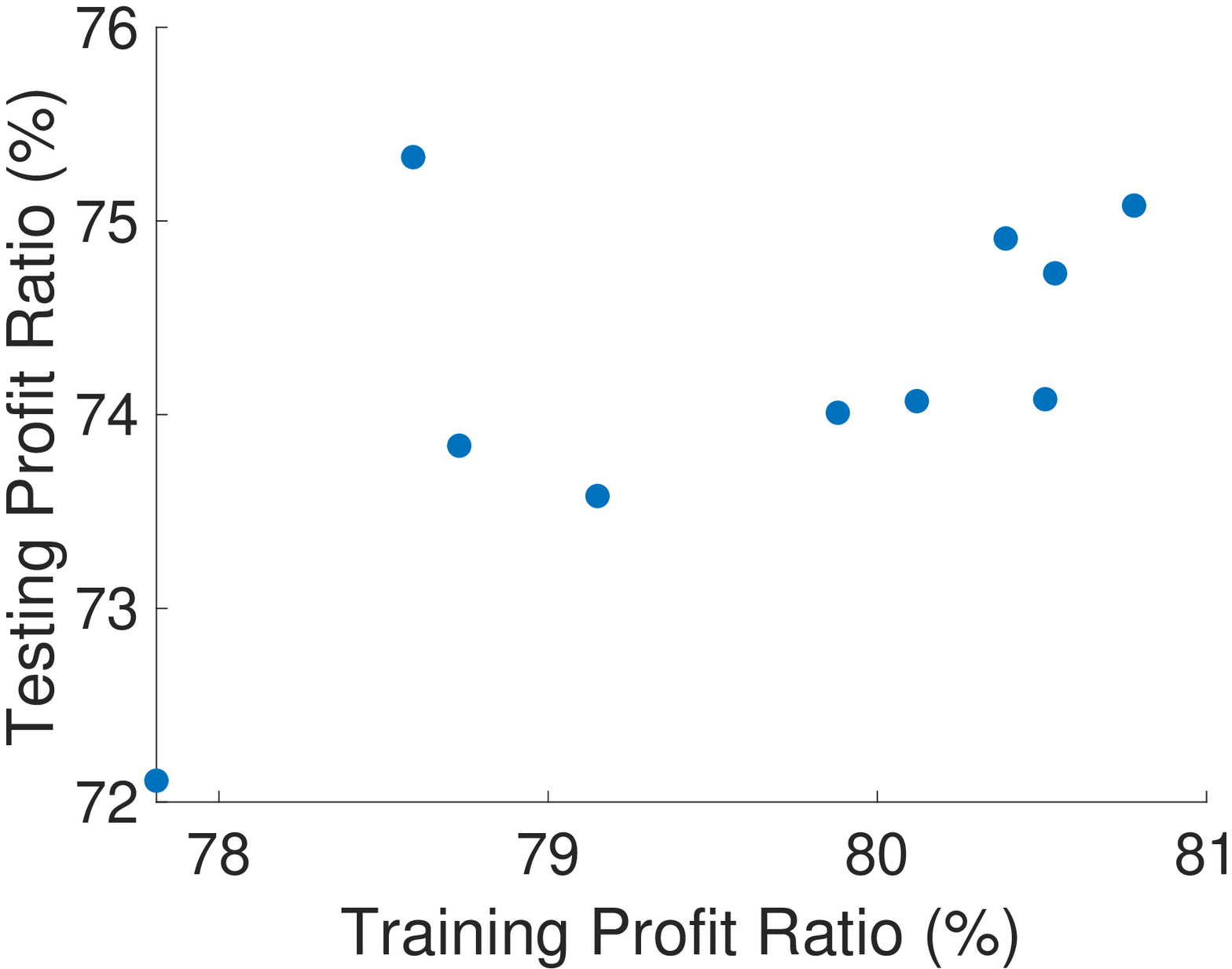}}
    \subfigure[WEST]{
		\includegraphics[trim = 0mm 0mm 10mm 0mm, clip, width = .45\columnwidth]{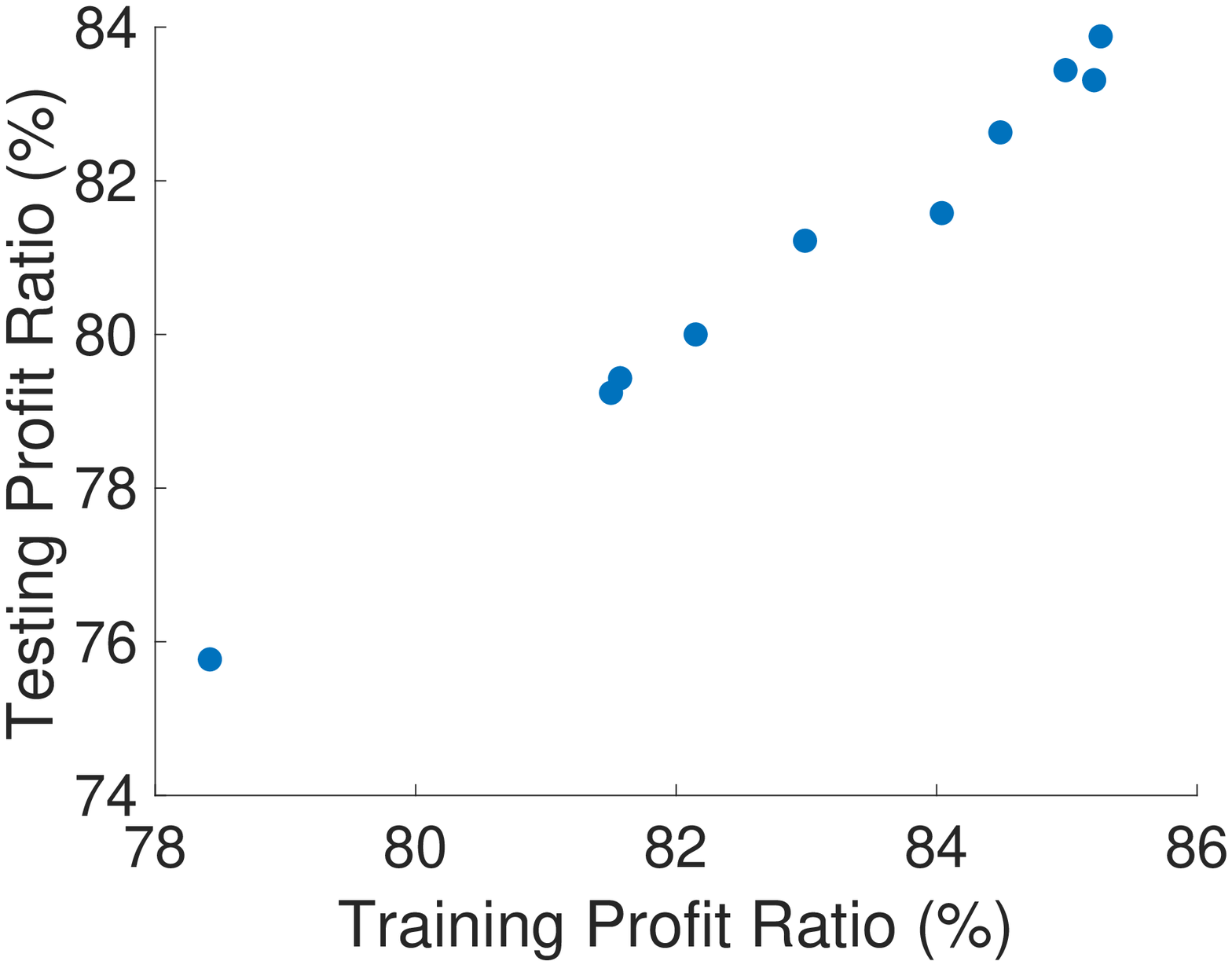}}
	\vspace{-0.4em}
	\caption{Training and testing profit relationship using MLP Setting 3.}
	\label{fig:train-test}
	\vspace{-0.6em}
\end{figure}

We compare the performance of MLP model settings for all zones in this subsection. First, we sequentially train 10 models with different random seeds for each location and setting. Then we select models with the best training arbitrage performance for testing arbitrage. Fig.~\ref{fig:train-test} shows the positive correlation between training profit and testing profit.  We use the profit ratio as the primary performance evaluation criteria to show how much profit the energy storage captured compared to the best possible market profit with perfect price prediction. 

{\footnotesize
\begin{table*}[ht]
\renewcommand{\arraystretch}{1.0}
\caption{Profit ratio, revenue and annual discharge of different energy storage duration and presumed marginal discharge cost (\$/MW)}
\centering
\vspace{-1.0em}
\resizebox{0.98\textwidth}{!}{
\begin{tabular}{lccccccccccccc}
\hline
\hline
\multirow{2}{*}{Zone}   & Duration & \multicolumn{4}{l}{Prorated Profit Ratio {[}\%{]}} & \multicolumn{4}{l}{Revenue {[}k\${]} / Annual Discharge {[}GWh{]}} & \multicolumn{4}{l}{Revenue per MWh discharged {[}\$/MWh{]}} \\
                        &     [Hours]               & 0MC         & 10MC       & 30MC       & 50MC       & 0MC             & 10MC           & 30MC           & 50MC           & 0MC          & 10MC         & 30MC          & 50MC          \\

\hline
\multirow{3}{*}{NYC}    & 1                    & 75.20 & 72.19 & 73.30 & 76.12
                        & 22.3/0.97       & 19.2/0.33      & 15.1/0.13      & 13.2/0.08      & 23.05        & 58.71        & 117.65        & 167.70        \\
                        & 2                  & 75.40 & 75.39 & 76.58 & 80.38
                        & 26.2/1.12       & 22.9/0.40      & 17.8/0.16       & 15.4/0.10       & 23.42        & 56.68        & 108.77        & 155.86        \\
                        & 4                 & 81.59 & 80.02 & 82.73 & 85.97
                        & 32.9/1.48        & 28.3/0.56       & 21.7/0.22       & 18.2/0.13       & 22.19        & 50.17        & 97.57         & 143.04        \\

\hline
\multirow{3}{*}{LONGIL} & 1                    & 71.15 & 68.99 & 64.28 & 65.28
                        & 35.8/1.23       & 32.7/0.51      & 25.3/0.21      & 22.4/0.13      & 29.13        & 63.60        & 122.42        & 168.32        \\
                        & 2                  &75.18 & 72.31 & 67.69 & 67.12
                        & 45.0/1.44       & 40.4/0.64      & 31.0/0.27      & 26.0/0.16      & 31.21        & 62.70        & 116.41        & 166.08        \\
                        & 4                 &79.22 & 76.83 & 72.68 & 73.53
                        & 55.1/1.81       & 49.6/0.85      & 37.8/0.35       &  32.0/0.21       & 30.47        & 58.12        & 108.79        & 154.95        \\

\hline
\multirow{3}{*}{NORTH}  & 1                    & 71.28 & 71.20 & 73.30 & 77.94
                        & 22.6/1.30       & 20.4/0.48      & 14.8/0.14      & 12.1/0.06      & 17.35        & 42.91        & 102.59        & 194.37        \\
                        & 2                  & 73.52 & 75.08 & 80.95 & 82.80
                        & 27.5/1.51       & 24.8/0.58      & 17.9/0.18       & 14.0/0.08       & 18.24        & 42.40        & 98.26        & 174.96        \\
                        & 4                 & 78.52 & 78.95 & 85.45 & 87.73
                         & 33.9/1.85        & 29.6/0.74       & 20.7/0.23       & 15.8/0.10       & 18.39        & 40.21        & 90.70         & 161.15        \\

\hline
\multirow{3}{*}{WEST}   & 1                    & 80.44 & 81.09 & 79.88 & 81.18
                        & 44.4/1.34       & 42.0/0.66      & 34.3/0.32      & 30.6/0.21      & 33.17        & 63.33        & 107.59        & 145.82        \\
                        & 2                  & 81.96 & 83.87 & 83.73 & 86.00
                        & 51.3/1.55       & 48.8/0.79      & 40.5/0.40      & 34.7/0.24      & 33.16        & 61.47        & 102.01        & 142.25        \\
                        & 4                 & 85.16 & 85.82 & 86.74 &88.03
                        & 60.3/1.89       & 56.3/0.98      & 46.5/0.49      & 38.2/0.28       & 31.87        & 57.23        & 94.66         & 135.78       \\
                        
\hline
\hline
\end{tabular}
}
\label{tab:profit}
\vspace{-2.5em}
\end{table*}}

\begin{figure}[ht]
\vspace{-1em}
\centering
\includegraphics[trim = 0mm 0mm 0mm 5mm, clip, width = .85\columnwidth]{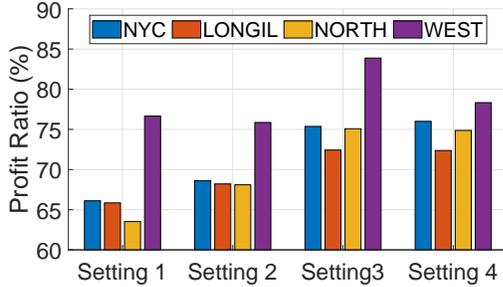}
\vspace{-1em}
\caption{Performance of different MLP model settings}
\label{fig: ratio}
\vspace{-1em}
\end{figure}

 The arbitrage performances using 2019 testing data are shown in Fig.~\ref{fig: ratio}. Comparing Setting 1 and 2 shows that using a longer RTPs look-back window slightly increases algorithm performance in most locations. Including DAPs as training data significantly improve the performances in Setting 3 and 4. DAP provides essential  information for real-time opportunity value prediction. However, a longer RTPs look-back window does not improve the performance of models with DAP, as shown in Setting 3 outperforms Setting 4 in most zones, so further tuning is needed in future work to understand the role of these extra features. 
 MLP model with Setting 3 has the best overall performance and lower computational cost. Therefore, we use this model setting for other simulations in this paper.

\subsection{Benchmarking with SDP and RL}
\begin{figure}[ht]
\vspace{-1em}
\centering
\includegraphics[width=0.9\columnwidth]{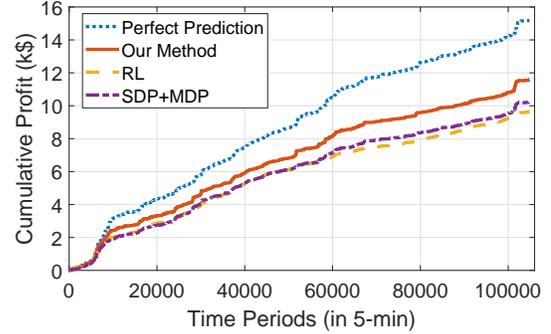}
\vspace{-1em}
\caption{Cumulative profits comparison of perfect price prediction arbitrage, RL, SDP, and our method in NYC. Energy storage setting: $E = 1$~MWh, $P = 0.5$~MW, $\eta^p=\eta^d=100\%$, and $c=\$10$/MWh.
}
\label{fig:RL}
\vspace{-1em}
\end{figure}

\rev{We further compare our method with an SDP benchmark 
and an RL benchmark with Q-learning \cite{wang2018energy}. The SDP benchmark predicts opportunity value function using a Markov process~\cite{zheng2022arbitraging}. In RL, we have 11 actions, 103 price states, and 121 SoC states, which take more than 1 hour to train for 5-min resolution arbitrage. Although prediction model training is offline, lower computational cost eases the model tuning process. Incorporating efficiency in SoC evolution constraints in RL requires near-continuous SoC state space. Therefore, we assume energy storage has 100\% efficiency in this simulation. Fig.~\ref{fig:RL} shows that our method significantly outperforms the RL benchmark with lower computational cost in an example price zone. Depending on different price zones, our proposed method improves profitability by 4-13\% 
and 18-30\% compared to SDP and RL, respectively. The profit ratio compared to arbitrage with perfect price prediction is shown in Fig.~\ref{fig: ratio}.}

\subsection{Model Transferability Comparison}

\begin{figure}[ht]
\vspace{-0.2em}
\centering
\includegraphics[width=0.8\columnwidth]{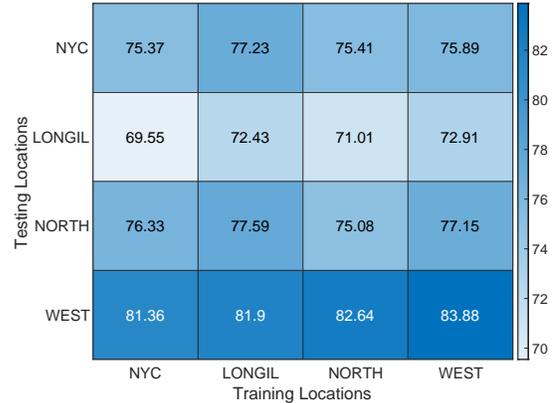}
\vspace{-1em}
\caption{Comparison of transferability for models trained at different locations.}
\vspace{-0.6em}
\label{fig:locations}
\end{figure}

We investigate model transferability using models trained using historical price data in one location to predict value functions at other locations. Fig.~\ref{fig:locations} shows the profit capabilities of models trained in different locations.

We find that the performance of value function prediction models trained from LONGIL and WEST perform superior to those trained from NYC and NORTH. These models are potentially more capable of predicting abnormal prices, which leads to greater overall profitability. Models with better performance in one location have good transferability and even outperform models trained by the original location price data. 

\subsection{Sensitivity Analysis of Energy Storage Parameters}

The effects of different energy storage parameters on 1~MW power rating storage arbitrage results are shown in Table.~\ref{fig: ratio}. We implement training and testing processes with three duration settings (1, 2, 4 hours) and four marginal discharge cost settings (0, 10, 30, and 50) in all locations. The arbitrage performance improves with a longer energy storage duration. With a longer energy storage duration, the energy storage is more likely to have spare energy capacity when a new charge/discharge opportunity happens. Increased marginal discharge cost leads to conservative operational decisions and higher unit revenue. In terms of zones, the storage achieved stable performance around 75\% to 90\% profit ratio except in LONGIL. Revenue-wise, the storage achieves the highest revenue in WEST followed by LONGIL, while NYC and NORTH provide similar revenue around 50\% of WEST. 

\vspace{-0.3em}


\section{Conclusion}\label{sec:conclusion}
\vspace{-0.2em}
In this paper, we proposed a deep-learning framework for energy storage price arbitrage. Instead of directly predicting real-time price, we use MLP to predict value functions of energy storage SoC levels. The proposed method is capable of capturing 65-90\% of the maximum possible profit in NYISO real-time market energy arbitrage, depending on energy storage characteristics and locations, which is superior to previous model-based and learning-based methods. While guaranteeing the algorithm's performance with regard to profitability, the algorithm is also light-weighted and can be trained and implemented with minimal computational cost. In this work, we only implement the standard MLP model for the learning process. The performance of the proposed method can be further improved with advanced deep learning structures and fine model tuning. \rev{The high computational efficiency of our algorithm also enables designing strategical bids except for price response. Our future research will use the proposed method for designing strategic bids and further explore market designs for energy storage and renewable integration.}

\bibliographystyle{IEEEtran}	
\bibliography{IEEEabrv,main}		
\end{document}